\documentclass[proof]{WileyASNA-v1}

\articletype{Article Type}%

\received{xx April 2022}
\revised{xx June 2022}
\accepted{xx June 2022}

\begin{document}

\title{An enigma of Przybylski's star: is there promethium on its surface?}

\author[1,2,3]{Sergei M. Andrievsky*}
\author[4]{Sergey A. Korotin}
\author[2]{Klaus Werner}
\author[1,2]{Valery V. Kovtyukh}

\authormark{Sergei M. Andrievsky et al.} 

\address[1]{\orgdiv{Astronomical Observatory}, \orgname{Odessa National University of the Ministry of 
Education and Science of Ukraine}, \orgaddress{\state{Shevchenko Park, 65014, Odessa}, \country{Ukraine}}}

\address[2]{\orgdiv{Institut f\"{u}r Astronomie und Astrophysik}, \orgname{Kepler Center for 
Astro and Particle Physics, Universit\"{a}t T\"{u}bingen}, \orgaddress{\state{Sand 1, 72076 T\"{u}bingen}, 
\country{Germany}}}

\address[3]{\orgdiv{GEPI}, \orgname{Observatoire de Paris, Universit\'e PSL, CNRS}, 
\orgaddress{\state{ 5 Place Jules Janssen, F-92190 Meudon}, \country{France}}}

\address[4] {\orgdiv{Physics of stars department}, \orgname{Crimean Astrophysical Observatory}, 
\orgaddress{\state{Nauchny 298409}, \country{Crimea}}}

\corres{*S.~M.~Andrievsky, Astronomical Observatory, Odessa National University of the Ministry of 
Education and Science of Ukraine, Shevchenko Park, 65014, Odessa, Ukraine \email{andrievskii@ukr.net}}

\abstract{We carried out a new attempt to check for the presence promethium lines in the spectrum of HD~101065 (Przybylski's star).
The neutron capture element promethium does not have stable isotopes and the maximum half-life time is about 18 years. 
Thus its presence in this peculiar star would indicate an ongoing process of irradiation of its surface layers with free 
neutrons. Unfortunately, almost all promethium lines are heavily blended with lines of other neutron capture elements and other 
species. We selected and analysed three lines of promethium (Pm~I and Pm~II) and came to the conclusion that at present
it is impossible to definitely claim the presence of this element in Przybylski's star atmosphere.}


\maketitle

\section{Introduction}

\cite{Andrievsky2022} described the peculiarities of the star HD~101065 (hereafter Przybylski's star) based
on the literature published since the discovery of this mysterious object. Here we only recall the most peculiar 
features that make this star one of the most unique objects of our Galaxy. 

\cite{Przybylski1961} and \cite{KronGord1961} for the first time noted the enormously strong 
line blanketing in the spectrum of Przybylski's star. \cite{Cowleyetal1977} did the first attempt 
to identify lines in the stellar spectrum. The authors identified many lines of lanthanides, indicating 
an extremely high abundance of heavy elements. Later this was confirmed by \cite{Cowleyetal2000} and 
\cite{Shulyaketal2010}.

For many years it was not possible to explain the very high abundance of the neutron capture elements
(which initially followed from the extremely peculiar spectrum of this star). There have been several attempts 
to find the source of the high-energy processes that could lead to such an extreme chemical peculiarity.
\cite{Cowleyetal2004} suggested that the reason for this is the mechanism of flare activity on the star.
\cite{Goriely2007} assumed that strong magnetic fields accelerate charged particles (protons and
$\alpha$-particles) that can cause spallation reactions, which alter the surface abundance of the
atmosphere gas. However, the magnetic field in the Przybylski star is not very strong (500--1500~G, according 
to \citealt{Hubrigetal2018}). 

Another possible problem with the protons and $\alpha$-particles as precursors of neutrons in spallation reactions 
could be their very fast thermalisation by elastic interactions with hydrogen nuclei of the atmosphere gas. 
After that, the thermalised protons and $\alpha$-particles are no longer capable to initiate spallation reactions.
 
\cite{Gopkaetal2007} and \cite{Gopkaetal2008} proposed an alternative explanation for the discussed phenomenon.
They suggested that Przybylski's star is a binary system with the second component being a neutron 
star, which is the source of high-energy particles that cause the abundance anomalies.
Although very fruitful, this approach was somewhat unrealistic in terms of the assumed geometry 
of the system and the reactions leading to the formation of neutron capture elements.  
 
\cite{Andrievsky2022} proposed a somewhat different hypothesis explaining several properties
of Przybylski's star. In particular, it was shown that the extremely high abundance of the 
neutron capture elements ($s$-process elements, in particular) can be explained by reactions 
between seed nuclei and thermalised photo-neutrons. The latter are formed in reactions between 
$\gamma$-quanta and the nuclei of some abundant elements, such as helium. The source of the 
$\gamma$-radiation could be a close neutron star companion of Przybylski's star.

\cite{AndrKovt2023} showed that despite the expectation of the presence of deuterium
in the atmosphere of Przybylski's star along with other neutron capture (heavy) elements, 
no deuterium lines were found. An explanation of this fact was given in terms of deuterium 
destruction by envelope convection.

Another interesting question concerning the peculiarities of Przybylski's star might arise. 
Is there evidence for the presence of unstable isotopes in the atmosphere of this star, which 
may indicate that the irradiation of its atmosphere by $\gamma$-quanta is an ongoing process? 
One such unstable element is promethium (Pm, Z = 61). It has only three "long-lived" isotopes 
with a half-life time of about 17.7, 2.6 and 5.5 years. The corresponding nuclei of $^{145}$Pm, 
$^{147}$Pm and $^{146}$Pm can be produced in the $s$-process in the following reactions:

$^{144}$Sm + $n$ = $^{145}$Sm

$^{145}$Sm + $e^{-}$ = $^{145}$Pm + $\nu$,

$^{146}$Nd + $n$ = $^{147}$Nd

$^{147}$Nd = $^{147}$Pm + $e^{-}$+ $\widetilde{\nu}$,

$^{145}$Pm + $n$ = $^{146}$Pm,

(if the lifetime of $^{145}$Pm is long enough to capture a neutron).


\cite{Cowleyetal2004} (see also  web-page of Dr. C. Cowley http://www-personal.umich.edu/$\sim$cowley/prz2r.html) 
claimed the detection of promethium (Pm~I and Pr~II) lines and technetium (Tc~I and Tc~II) 
lines in the spectrum of Przybylski's star. The first attempt to calculate Pm~II line radiative characteristics (46 lines) 
was made by \cite{Fivetetal2007}. Unfortunately, the promethium lines are usually strongly blended, in particular, 
with the lines of neutron capture elements (Ce~II, Pr~II, Nd~II, Sm~II and others), which are extremely 
overabundant. From the figures provided by \cite{Fivetetal2007}, it can be concluded that the detections 
of Pm lines are not entirely certain, and the authors indicate this fact in their Sect. 6. 
At the same time, the authors claim that the most reliable line identification could be Pm~II $\lambda$ 5576.02 \AA. 
Nevertheless, the authors did not consider together with the fairly strong blending line of Fe~I ($\lambda$ 5576.09 \AA), 
a Co~I blend consisting of about thirty components. According to our calculations, these lines can dominate 
the blend. This generally makes the classification of the $\lambda$ 5576 \AA~ blend as a promethium-dominated 
mixture very uncertain. (According to $priv.~comm.$ of our refree another blending line of Nd~II
at 5575.992 \AA\ may have a significant contribution to the discussed blend.
Another line at $\lambda$ 5546.08 \AA, marked by C. Cowley as Pm~II line, is actually heavily blended with a Y~II line.

A very interesting consideration of the promethium problem was made by \cite{Ryabetal2008}. The author 
noted that the lines, previously identified as Pm~II lines, do not match the observations. For example, 
two lines 6659.07 \AA~ and 6772.28 \AA~ identified by \cite{Cowleyetal2004} as Pr~II lines may actually be 
well reproduced by Ce~II lines. It is one of the rare-earth elements with an extremely high abundance
in the Przybylski's star atmosphere. In this paper, we report on our new attempt to search for promethium 
lines in Przybylski's star. 
    
\section{Synthetic spectrum calculation}

The principal problem in determining the promethium contribution to a particular
blend is the high abundance of neutron capture elements, which significantly complicates 
reliable identification.

We selected only a few Pm~I and Pm~II lines from the NIST database (https://www.nist.gov/) 
for which the lower and upper level excitation potentials were determined.  After preliminary 
selection by synthetic spectrum test calculations, we settled on choosing two Pm~I lines, namely 
$\lambda$ 4959.46 \AA~and 5058.31 \AA, and one Pm~II line at 5576.02 \AA. The rest of the Pm~I and Pm~II 
lines are heavily blended and practically cannot be used for analysis. 

To generate the synthetic spectrum, we used the SynthV code of \cite{Tsymbaletal2019} 
and two atmosphere models. The calculations were carried out in local thermodynamical equilibrium (LTE) 
approximation. The first model is the standard model from \cite{CastKur2003} with solar elemental
abundances and the second one is the LL models from \cite{Shulyaketal2004}. The LL models code allows to 
take into account the line opacity, considering the abundances of individual elements and the 
possible stratification of several chemical elements. To calculate the LL models we used the atmospheric parameters
of Przybylski's star ($T_{\rm eff}=6400$\,K, $\log g=4.2$), its chemical composition and the stratification of Si, Ca, Fe, 
and Ba, following \cite{Shulyaketal2010}. According to this work, Przybylski's star shows an overabundance of the neutron 
capture elements by 2--4 orders of magnitude compared to solar abundances. Such overabundances drastically
change the structure of the atmosphere.

\begin{figure} 
\resizebox{\hsize}{!}{\includegraphics{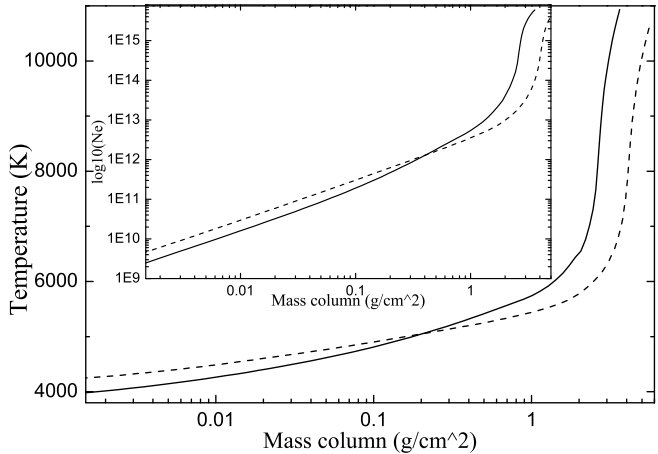}}    
\caption{Temperature and electron density structures in the model atmospheres for Przybylski's star. 
The dashed line is for the standard atmosphere model and the solid line is for the LL models}
\label{Ne&Temp}
\end{figure}

Fig. \ref{Ne&Temp} shows the significant difference in the temperature and electron density stratifications 
between the LL models (including neutron capture element overabundances and element stratifications) and the
standard solar-abundance and chemically homogeneous atmosphere model. The temperature in deep atmospheric 
layers is much hotter in the LL models due to enhanced line blanketing, while in the uppermost layers the 
situation is reversed because of stronger line cooling. A similar effect concerns the electron density 
distribution. It should be noted that at the same time a shift of the radiative energy flux from the ultraviolet 
region to the infrared occurs. This is a consequence of the large number of the lines of the neutron capture 
elements in the ultraviolet. A detailed description of changes in the atmosphere structure is 
given in \cite{Shulyaketal2010}.

\begin{figure} 
\resizebox{\hsize}{!}{\includegraphics{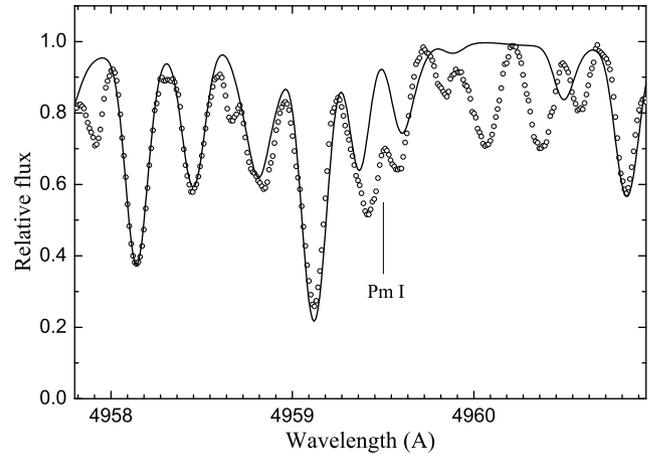}}    
\caption{Details of the observed spectrum of Przybylski's star (open circles), covering
the possible Pm~I 4959.46 \AA~line and the synthetic spectrum (solid line) without the calculated Pm line
profile. The position of the promethium line is indicated by the vertical bar}
\label{4959_Pm}
\end{figure}

\begin{figure} 
\resizebox{\hsize}{!}{\includegraphics{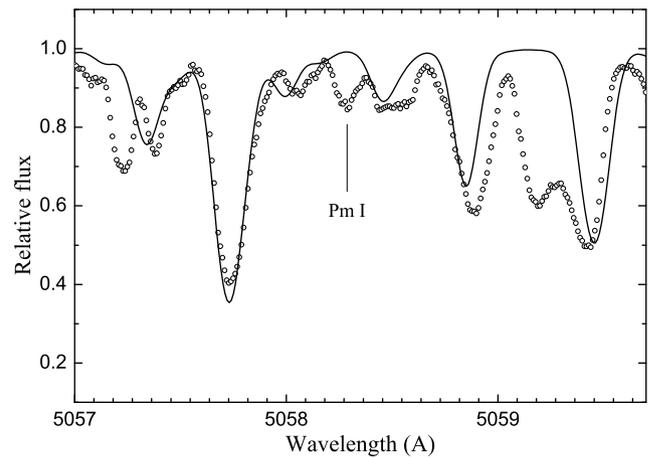}}    
\caption{Same as Fig. \ref{4959_Pm} but for the possible Pm~I 5058.31 \AA~line}
\label{5058_Pm}
\end{figure}

\begin{figure} 
\resizebox{\hsize}{!}{\includegraphics{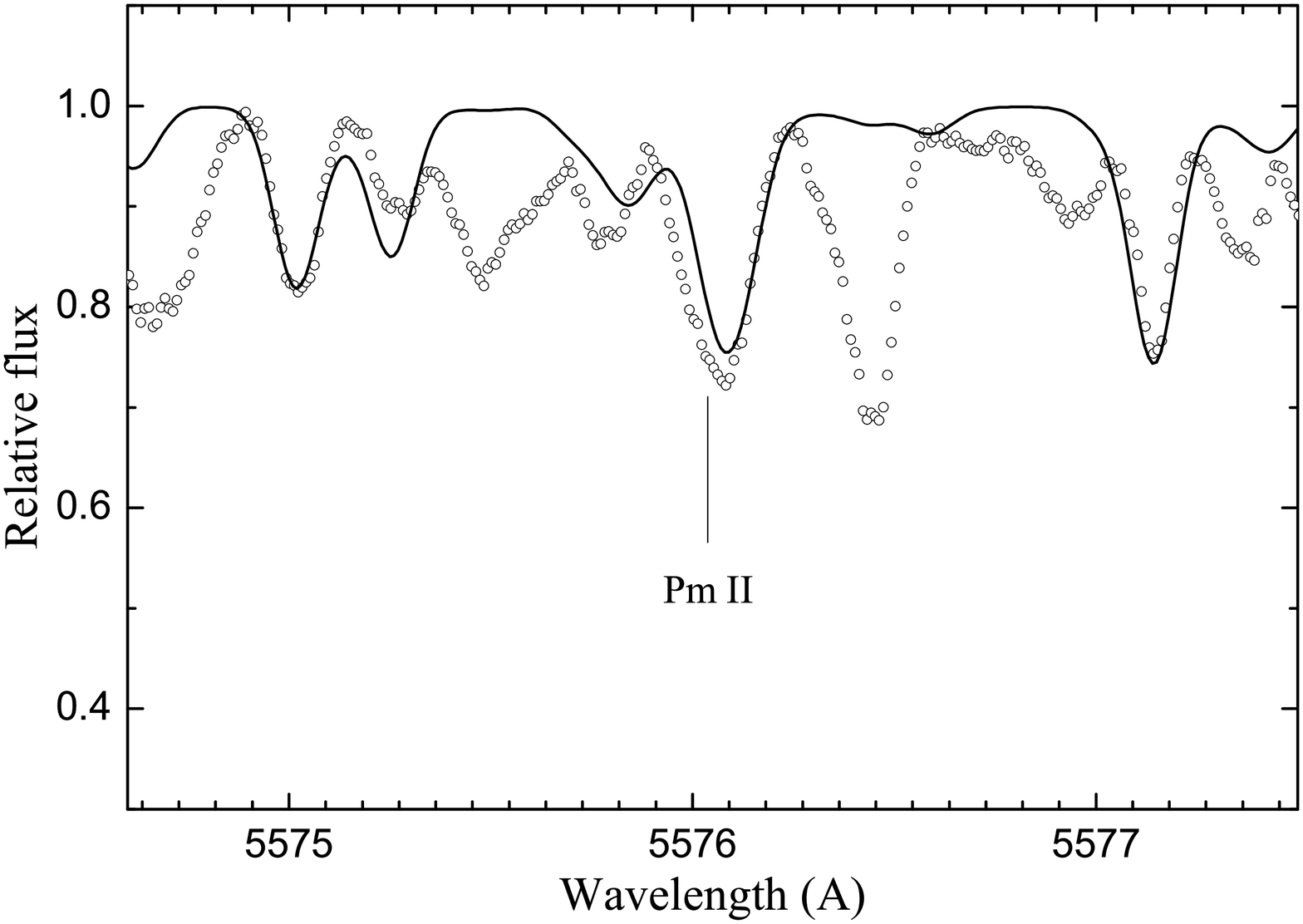}}    
\caption{Same as Fig. \ref{4959_Pm} but for the possible Pm~II 5576.02 \AA~line}
\label{5576_Pm}
\end{figure}

Figs. \ref{4959_Pm}, \ref{5058_Pm}, and \ref{5576_Pm} show the results of our synthetic spectrum calculations 
together with the observed spectrum in the vicinity of the above mentioned lines. The observed spectrum 
has been retrieved from the ESO Archive (HARPS, S/N = 140, 2005-04-03, 05:07:42 observation).

\subsection{Detailed spectrum synthesis of the Pm~II 5576.02 \AA~line}

\begin{figure} 
\resizebox{\hsize}{!}{\includegraphics{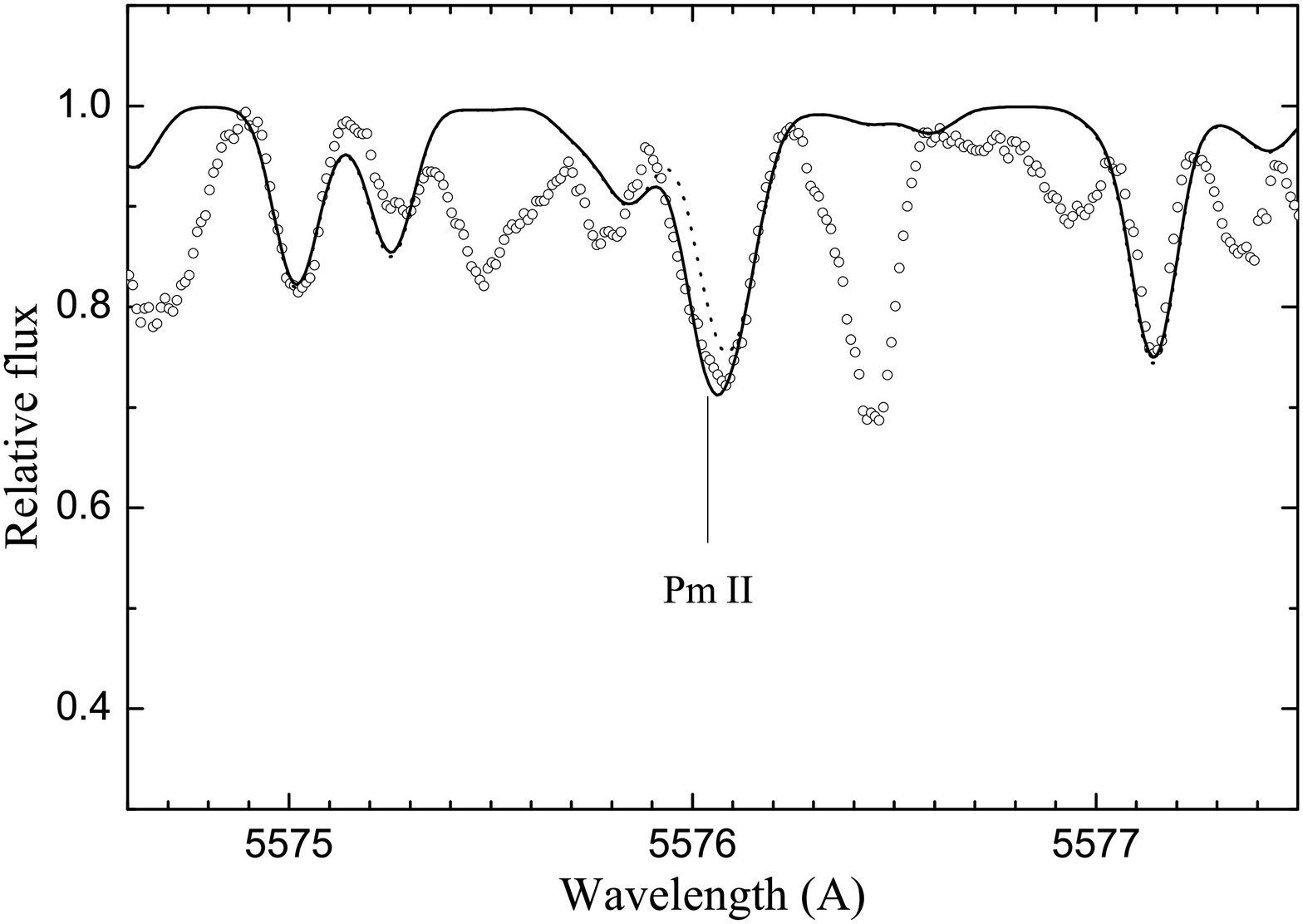}}    
\caption{Observed and synthetic spectra in the vicinity of the Pm~II 5576.02 \AA~line calculated with the 
oscillator strength from \cite{Fivetetal2007}. Observed spectrum -- open circles, dashed line -- promethium is absent, 
solid line -- promethium abundance is $\log\epsilon$(Pm/H) = +1.8}
\label{5576_PmII_Synth}
\end{figure}

We used the partition functions and the value of $\log~gf$ for the line Pm~II 5576.02 \AA~calculated by \cite{Fivetetal2007}
in order to obtain a quantitative estimate of the possible content of promethium in Przybylski's star. Our synthetic profile 
calculations showed that the best agreement with the observed profile is achieved if we adopt an abundance
$\log\epsilon$(Pm/H) = +1.8. Fig.~\ref{5576_PmII_Synth} shows the resulting best fit. (Note that as stated above, this blend 
may also be contributed by the line(s) of the rare-earth element, like Nd). 

\section{Discussion and conclusion}

The spectrum of Przybylski's star is very complex, as can be seen even from the three fragments of the 
spectrum in Figs. \ref{4959_Pm}-- \ref{5576_Pm}. Some lines cannot be identified at all and thus they are not described 
by spectrum synthesis. To generate a synthetic spectrum, we used the extended Vienna Atomic Line Database \citep{Ryabetal2015} 
and the results of abundance determinations from \cite{Shulyaketal2010}. These abundance values, however, are not highly accurate. 
For instance, the abundances of certain elements in different ionisation stages deviate by about 1--2 orders of magnitude. 
Even for the close lines of a particular element in the same ionization stage, the observed and calculated profiles can differ 
drastically. Therefore, we adjusted the abundances of such elements, like Ce, Nd, and some others in order to match the 
respective observed and synthetic line profiles, at least in the plots presented here.

{\bf Pm~I~4959.46~\AA~line.} Numerical details of the synthetic spectrum calculation in the vicinity
of this line are presented in Table~\ref{Line1}. The tungsten line at 4959.369 \AA~ heavily blends the possible Pm~I line. 
We reduced the abundance of this element, derived by \cite{Shulyaketal2010}, by 0.1 dex, otherwise this line
turns out to be very strong, its profile is deeper than the observed profile of the blend, and the W~II line core is
blueshifted compared to the blend core. The Ce~II line  4959.60 \AA~ can also make a significant contribution
to the overall blend.

\begin{table}
\caption{Wavelengths, identifications, and residual fluxes of some spectral lines in the vicinity of the Pm~I 4959.46~\AA~line}
\label{Line1}
\begin{tabular}{lcc}
\hline   
 4959.0762& Tm II&  0.9922\\
 4959.1124& Ce II&  0.9700\\
 4959.1190& Nd II&  0.0721\\
 4959.1977& Fe I &  0.9478\\
 4959.3690& W II &  0.3918\\
 4959.46  & Pm I &        \\
 4959.5306& Cr I &  0.9942\\
 4959.5992& Ce II&  0.6188\\
 4959.6650& Co I &  0.9810\\
 4959.6729& Co I &  0.9800\\
 4959.6750& Co I &  0.9932\\
 4959.6831& Co I &  0.9856\\
 4959.6852& Co I &  0.9800\\
 4959.6974& Co I &  0.9589\\
 4959.8710& Ru I &  0.9686\\
 4959.8970& Ce II&  0.9681\\
 4960.1980& Eu I &  0.9940\\
 \hline
\end{tabular}
\end{table}

    {\bf Pm~I~5058.31~\AA~line.} At this wavelength, the synthetic spectrum does not show
    any strong lines from other elements, however, an unambiguous conclusion about the presence 
    of this Pm line cannot be made, given the fact that even in this spectrum fragment we note 
    observed lines that cannot be identified.

\begin{table}
\caption{Same as Table \ref{Line1} but for the Pm~I line 5058.31 \AA}
\label{Line2}
\begin{tabular}{lcc}
\hline
 5057.9920& Nb I &  0.7962\\
 5058.0316& Fe I &  0.9755\\
 5058.1600& Hf II&  0.9315\\
 5058.31  & Pm I &        \\
 5058.3111& Ce II&  0.9931\\
 5058.4465& Ce II&  0.9629\\
 5058.4521& Ce II&  0.7947\\
 5058.4961& Fe I &  0.9792\\
 5058.5500& Re I &  0.9359\\
 5058.6909& Yb II&  0.9939\\
 5058.8500& Sm II&  0.3898\\
 5059.0186& Tm II&  0.9901\\
 5059.3510& Nb I &  0.9884\\ 
 \hline                
\end{tabular}          
\end{table}

    {\bf Pm~II~5576.02~\AA~line.} This Pm~II line is blended with many components of a Co~I line. 
    Another (though quite distant) contributor is the Fe~I line with a relatively small residual flux.

\begin{table}
\caption{Same as Table \ref{Line1} but for the Pm~II line 5576.02 \AA}
\label{Line3}
\begin{tabular}{lcc}
\hline
 5575.7090& Yb II&  0.9635\\
 5575.7693& Co I&  0.9946\\
 5575.7909& Co I&  0.9913\\
 5575.8104& Co I&  0.9900\\
 5575.8171& Co I&  0.9583\\
 5575.8280& Co I&  0.9901\\
 5575.8317& Co I&  0.9724\\
 5575.8434& Co I&  0.9915\\
 5575.8444& Co I&  0.9831\\
 5575.8551& Co I&  0.9907\\
 5575.8567& Co I&  0.9938\\
 5575.8795& Co I&  0.9946\\
 5575.8853& Co I&  0.9913\\
 5575.8882& Co I&  0.9938\\
 5575.8891& Co I&  0.9900\\
 5575.8906& Co I&  0.9915\\
 5575.8909& Co I&  0.9901\\
 5575.9763& Co I&  0.9936\\
 5575.9968& Co I&  0.9903\\
 5576.0148& Co I&  0.9896\\
 5576.02  & Pm II&        \\
 5576.0305& Co I&  0.9910\\
 5576.0331& Co I&  0.9688\\
 5576.0441& Co I&  0.9838\\
 5576.0444& Co I&  0.9943\\
 5576.0526& Co I&  0.9936\\
 5576.0851& Co I&  0.9943\\
 5576.0888& Fe I&  0.7437\\
 5576.0917& Co I&  0.9910\\
 5576.0967& Co I&  0.9896\\
 5576.0999& Co I&  0.9903\\
 5576.1009& Co I&  0.9936\\
 5576.1570& Nb I&  0.9659\\
 5576.3190& W I &  0.9920\\
 5576.4400& Zr II& 0.9729\\
 5576.5904& Mn I&  0.9572\\
 \hline
\end{tabular}
\end{table}

Since this blend is significantly affected by the Co~I multicomponent feature and the Fe~I line, we cannot say with certainty that a promethium 
line contribution is really present in this blend (see Fig. \ref{5576_PmII_Synth}). Many factors can affect our conclusion. Among these are 
the uncertainty in the adopted model atmosphere structure, stratification of chemical elements (which is unknown for many species 
present in the atmosphere of this star), and the possible uncertainty in the calculated oscillator strength of the considered Pm~II line. If we formally 
use the Pm abundance derived from this line, and try to fit the observed profiles of two Pm~I lines (for which the oscillator strength were not determined), 
namely 4959.46 \AA~ and 5058.31 \AA, we get unrealistic results: $\log~gf\epsilon$ = 5.5 and 5.1 respectively. If we arbitrarily assume that 
$f$ is about 1 for each of two lines, then $\log~gf \approx 0.8-0.9$, because in the NIST database the total angular momentum $J$ value for the lower atomic 
levels of these lines is 5/2 and 7/2, respectively, and statistical weight is $g = 2J+1$. Therefore, we get $\log\epsilon$(Pm/H) $\approx 4.2-4.7$, 
and the difference in promethium abundance derived from the Pm~I lines and Pm~II line reaches more than two orders of magnitude. Although \cite{Shulyaketal2010} 
also obtain a large difference in abundances of some elements derived from the lines of two or three ionization stages, we question our obtained 
results on the promethium abundance in Przybylski's star. Significantly higher spectral resolution and high signal-to-noise ratio of the stellar spectrum, 
additional data on the stratification of neutron capture elements in the atmosphere, and more accurate data on the characteristics 
of the lines of rare-earth elements could help solve some of the problems of identifying promethium lines in the blends under discussion.
  
Summarizing up, one can state that at present it is impossible to draw an unambiguous conclusion about the presence of promethium 
in the atmosphere of Przybylski's star. 


\subsection*{ACKNOWLEDGEMENTS}

SMA and VVK are grateful to the Vector-Stiftung at Stuttgart, Germany, for support within the 
program "2022--Immediate help for Ukrainian refugee scientists" under grants P2022-0063 and P2022-0064. 
This work is based on data retrieved from the ESO Archive (P.I.: Hatzes A.P. Programme  075.C-0234).
We are very grateful to our referee for valuable comments that have improved this paper.

\subsection*{Conflict of interest}

The authors declare no potential conflict of interests.

\end{document}